\documentclass[usegraphicx,useAMS,usenatbib]{mn2e}

\newcommand{\beq}{\begin{equation}}
\newcommand{\eeq}{\end{equation}}
\newcommand{\beqa}{\begin{eqnarray}}
\newcommand{\eeqa}{\end{eqnarray}}

\def\eq#1{equation~(\ref{#1})}
\def\lexp{\mathop{{\bigl\langle}}\nolimits}
\def\rexp{\mathop{{\bigr\rangle}}\nolimits}
\def\half{{\frac12}}

\def\RAMSES{{\sc RAMSES}}

\def\G{{\cal G}}

\def\nbar{{\bar n}}
\def\Nbar{{\bar N}}
\def\xibar{{\bar \xi}}
\def\mubar{{\bar \mu}}
\def\bbar{{\bar b}}
\def\bbk{{\bar b}_k}

\def\d{\textrm{d}}
\def\min{{\rm min}}
\def\tot{{\rm tot}}
\def\red{{\rm red}}
\def\blue{{\rm blue}}

\def\thalf{{\textstyle{1 \over 2}}}

\def\Mpc{\, h^{-1} {\rm Mpc}}

\begin{document}

\title[Void Statistics and Hierarchical Scaling]
{Void Statistics and Hierarchical Scaling in the Halo Model}

\author[J. N. Fry and S. Colombi]
{J. N. Fry$^{1,2}$\thanks{E-mails: fry@phys.ufl.edu (JNF); colombi@iap.fr (SC)}
and S. Colombi$^{2\star}$ \\
$^1$Department of Physics, University of Florida, 
Gainesville FL 32611-8440, USA \\
$^2$Institut d'Astrophysique de Paris, CNRS UMR 7095 and UPMC, 
98bis bd Arago, F-75014 Paris, France }

\maketitle

\begin{abstract}

We study scaling behaviour of statistics of voids 
in the context of the halo model of nonlinear large-scale structure.
The halo model allows us to understand why the observed galaxy 
void probability obeys hierarchal scaling, even though the premise 
from which the scaling is derived is not satisfied.
We argue that the commonly observed negative binomial scaling 
is not fundamental, but merely the result of the specific 
values of bias and number density for typical galaxies.
The model implies quantitative relations between 
void statistics measured for two populations of galaxies, 
such as SDSS red and blue galaxies, and their number density and bias.

\end{abstract}


\section{Introduction}

Understanding the behaviour of voids in the galaxy distribution 
is one of the remaining unsolved problems of large-scale structure.
Voids are a powerful probe of nonlinear large scale structure.
They probe high order statistical properties, 
but do so on scales that should be accessible in perturbation theory.
One interesting property of voids is a scaling behaviour 
implied in the hierarchical model of higher order clustering.
The hierarchical scaling has been verified many times, in a variety 
of samples, including the CfA redshift survey \citep{MLR87,VGH94}, 
Perseus-Pisces \cite{FGHMS89},  
the Southern Sky Redshift Survey \citep{MSdC92}, 
and {\it IRAS} 1.2-Jy \citep{BSDFYH93}.
Particularly intriguing are recent results from 2dFGRS \citep{2dF04,CNGB07} 
and from DEEP2 and SDSS \citep{CCWNYCGDK05,TCNPW08}, 
in which the scaling continues to hold with improved precision over 
larger scales, for both magnitude selected subsamples and random dilutions.

However, in data \citep{BSDFYH93,GAPM94,Cetal04,RBM06,RBM07},  
and in numerical simulations \citep{FCFBST11}, 
the hierarchical clustering on which the scaling is based is not obeyed.
The hierarchical normalization removes much of the variation, 
but the hierarchical amplitudes still depend on scale, 
and the premise of the scaling does not hold in detail.
A recent alternative to purely hierarchical behaviour is 
provided by the halo model \citep{MF00a,MF00b,SSHJ01}. 
In this paper we show that void scaling can be understood 
in the halo model.

In Section 2, we review statistics and display the connection 
between voids and correlation functions, and we apply the halo model.
Many common models are realizations of the halo model, 
and we present several of these in Section 3.
In Section 4 we test the model against numerical results 
and present several halo model scaling relations.
Section 5 contains a final discussion.

\section{Statistics of Voids}

\subsection{Generating Functions}

The probability that a volume be empty of galaxies, or void, 
is an intriguing statistical measure, accessible to perturbation 
theory on large scales and yet 
an inherently nonlinear statistic on all scales.
We study the properties of voids in the context of the halo model, 
the essence of which is that galaxies come in groups or clusters 
embedded in haloes; the number of galaxies 
is then the sum over haloes of the number within each halo.
The generating function formulation of the halo model 
\citep{FCFBST11} is useful for studying combinatorics, 
and particularly voids.
Let the probability and moment generating functions be 
\beqa
G(z) &=& \sum_{n=1}^\infty P_n \, z^n = \lexp z^n \rexp , \\
M(t) &=& \sum_{k=1}^\infty {1 \over k!} \,  m_k \, t^k , 
\eeqa
where $ P_n $ is the probability that a randomly placed 
volume contains $n$ galaxies, and $ m_k $ is the order $k$ 
factorial moment of the distribution, 
$ m_k = \Nbar^k \, \mubar_k = \lexp n^{[k]}\rexp $, 
where $ n^{[k]} = n (n-1) \cdots (n-k+1) = n!/(n-k)! $.
Since both probabilities and moments can be obtained as 
derivatives of $G$, probabilities as 
\beq
P_n = {1 \over n!} \, {\d^n \over \d z^n} G(t) \Bigr|_{z=0} , 
\eeq
and moments as 
\beq
m_k = \lexp n (n-1) \cdots (n-k+1) \rexp 
= {\d^k \over \d z^k} G(z) \Bigr|_{z=1} , \label{m_k} 
\eeq
we see that the generator $ M(t) $ of moments, 
so that $ m_k = \d^k M(t)/\d t^k |_{t=0} $ 
is thus $ M(t) = G(t+1) $ \citep[cf.][]{SS93}.
Connected, irreducible, discreteness corrected moments 
$ k_n = \Nbar^n \xibar_n $ 
are similarly obtained from $ K(t) = \ln M(t) = \ln G(t+1) $.

In terms of the irreducible $ \xibar_n $, the probability 
that a volume $V$ be empty of galaxies, or void, 
is then a sum over all orders, 
\beq
P_0 = G(0) = \exp \left[ K(-1) \right] = \exp \Bigl[ \sum_{k=1}^\infty 
 {(-1)^{k} \over k!} \, \Nbar^k \xibar_k \Bigr] , \label{P0}
\eeq
a result also obtained by considering Venn diagrams and 
contour integrals in the complex plane \citep{FGJW76,W79,F86}.
From the void probability, we can write the statistic 
\beq
\chi = - {\ln P_0 \over \Nbar} =   \sum_{k=1}^\infty 
 {(-1)^{k} \over k!} \, \Nbar^{k-1} \xibar_k  . \label{chi}
\eeq
In observations, in perturbation theory, and in stable clustering,
we often take the volume-averaged correlations to follow the 
so-called hierarchical pattern, 
\beq
\xibar_n = S_n \, \xibar^{n-1} .  \label{S_n}
\eeq
($ S_2 = 1 $).
In the hierarchical case, the void probability becomes 
\beq
\chi = \sum_{k=1}^\infty {(-1)^{k-1} \over k!} \, S_k \, (\Nbar\xibar)^{k-1} 
= \chi(\Nbar\xibar) , \label{chi_hier}
\eeq
a power series in the variable $ \Nbar \xibar $.
This is the hierarchical scaling relation: 
the void statistic $ -\log P_0/\Nbar $ is a function of 
the scaling variable $ \Nbar \xibar $, 
where the void probability $ P_0 $, the mean count 
$ \Nbar = \langle N \rangle $, and the scaling variable 
$ \Nbar \xibar =  (\langle N^2 \rangle - \langle N \rangle^2 
 - \langle N \rangle) / \langle N \rangle $ 
are all observationally measurable quantities.
When $ \Nbar\xibar $ is small, $ \chi \to 1 $, the Poisson 
result $ P_0 = e^{-\Nbar} $, 
with clustering appearing only as a small correction.
When $ \Nbar\xibar $ is large, the void scaling behaviour 
is a strong test of hierarchical clustering to high orders.
A similar scaling behaviour has been found for gaps in the rapidity 
distribution resulting from proton-antiproton, 
proton-nucleus, and relativistic heavy ion collisions 
\citep{H92,M96,GDGCKAS01}.

Several models have been presented with specific analytic forms 
for the void scaling function $ \chi(\Nbar\xibar) $, 
useful against which to compare observational and numerical results.
Details are contained in Appendix~A.

Void scaling has been tested and found to hold in observational 
data from the CfA redshift survey \citep{MLR87,VGH94}, 
Perseus-Pisces \citep{FGHMS89}, 
the Southern Sky Redshift Survey \citep{MSdC92}, 
the {\it IRAS} 1.2-Jy redshift catalog \citep{BSDFYH93}, and more recently 
in the 2dFGRS \citep{2dF04,CNGB07}, and DEEP2 and SDSS \citep{CCWNYCGDK05}.
However, it is not clear that the scaling should be obeyed: 
although the normalization 
to $ S_k = \xibar_k/\xibar^{k-1} $ removes much of the dependence 
of $ \xibar_k $ on scale, the $ S_k $ are not in fact constant 
\citep{BSDFYH93,GAPM94,Cetal04,CNGB07,RBM06,RBM07}, 
and the galaxy distribution does not obey \eq{S_n}.
To understand this, is interesting to look at implications for 
voids in the halo model.

\subsection{The Halo Model}

Reduced to its most basic terms, in the halo model total galaxy count 
is the sum over clusters of the number of objects in a cluster.
On large scales, boundary effects are unimportant 
and clusters can be considered as point objects that 
are either entirely inside or entirely outside, 
the point cluster limit of the halo model \citep{FCFBST11}.
In the limit that clusters are unresolved (the point cluster limit, 
each cluster is either entirely within $V$ or entirely outside of $V$), 
and all clusters have identical occupation distribution (each cluster has 
the same mean count $ \Nbar_i $ and higher order moments $ \mubar_{n,i}$), 
the generating function total count probabilities is 
the composition of the halo number and halo occupancy 
generating functions, 
\beq
G(z) = g_h \left[ g_i \left( z \right) \right] . \label{Gz}
\eeq
and galaxy count moments are simply related to correlations 
$ \xibar_{k,h} $ of halo number 
and moments $ \mubar_k $ of halo occupation, 
with $ \Nbar_g = \Nbar_h \Nbar_i $, and 
\beqa 
\xibar_2 = \xibar_{2,h} &+& {\mubar_{2,i} \over \Nbar_h} \label{xi2} \\
\xibar_3 = \xibar_{3,h} &+& {3\mubar_{2,i} \xibar_{2,h} \over \Nbar_h} 
 + {\mubar_{3,i} \over \Nbar_h^2} \label{xi3} \\ 
\xibar_4 = \xibar_{4,h} &+& {6\mubar_{2,i} \xibar_{3,h} \over \Nbar_h} 
 + {(4 \mubar_{3,i} + 3 \mubar_{2,i}^2) \xibar_h  \over \Nbar_h^2} 
 + {\mubar_{4,i} \over \Nbar_h^3}  \label{xi4} \\
\xibar_5 = \xibar_{5,h} &+& {10 \mubar_{2,i} \xibar_{4,h} \over \Nbar_h} 
 + {(10 \mubar_{3,i} + 15 \mubar_{2,i}^2) \xibar_{3,h} \over \Nbar_h^2}
 \nonumber \\
&+& {(10 \mubar_{2,i} \mubar_{3,i} + 5\mubar_{4,i}) \xibar_{2,h} 
\over \Nbar_h^3} + {\mubar_{5,i} \over \Nbar_h^4} , \label{xi5}
\eeqa
The moments $ \xibar_k $ are in general the sum of many terms,  
with different dependences on scale, and galaxies do not in general 
have the constant amplitudes of the hierarchical scaling pattern.
However, if only the underlying cluster correlations obey 
$ \xibar_{k,h} = S_k \, \xibar_h^{k-1} $, 
these relations contain a more subtle scaling.

The combinatorics implied by the composition of generating functions 
in \eq{Gz} and the general pattern of equations (\ref{xi2})--(\ref{xi5}) 
remain true in the full halo model, in which 
occupation statistics depend on halo mass, with a distribution 
described by the halo mass function $ \d n/\d m $ and in which haloes 
can span the boundaries of $V$, with two modifications \citep{FCFBST11}.
First, when haloes are not identical but range over a distribution 
of masses, every term in $ G(z) = \lexp z^N \rexp $, 
and in particular the occupation moments $ \mubar_{n,i} $,  
are further averaged over the halo mass function.
After averaging over haloes of different mass, with mass-dependent 
occupation distribution and correlation strength, the net effect 
is to replace the occupation moment $ \mubar_k $ with 
\beq
\mubar_k \to {\bbk \over \bbar} \, \mubar_k \label{bmu_k}
\eeq
and halo correlations with 
\beq
\xibar_{k,h} \to {\bbar^k \over \bbar_h^k} \, \xibar_h \label{bxi_k}
\eeq
where the mean halo bias $ \bbar_h $ is $ b(m)$ as given 
in \citet{MJW97}, averaged over the occupied halo mass function, 
\beq
\bbar_h = { \int \d m \, (\d n/\d m) \, b(m) 
 \over \int \d m \, (\d n/\d m) } ,  \label{b_h}
\eeq
the mean galaxy bias $\bbar$ is weighted by occupation number, 
\beq
\bbar = { \int \d m \, (\d n/\d m) \, b(m) \, \langle N \rangle 
 \over \int \d m \, (\d n/\d m) \, \langle N \rangle} ,  \label{bb}
\eeq
and $ \bbk $ is weighted by the occupation number 
factorial moment $ \langle N^{[k]} \rangle $, 
\beq
\bbk = { \int \d m \, (\d n/\d m) \, \langle N^{[k]} \rangle \, b(m)
\over  \int \d m \, (\d n/\d m) \, \langle N^{[k]} \rangle } , 
\label{b_k}
\eeq
($ \bbar = \bbar_1 $).
The weighted bias factors $ \bbar_k $ are generally of order unity 
(but since higher order bias factors are weighted by higher powers of mass 
and bias is typically an increasing function of mass, $ \bbar_k $ 
is rising with $k$).
If halo correlations are hierarchical, 
$ \xibar_{k,h} = S_{k,h} \, \xibar_{2,h}^{k-1} $, galaxy correlations 
are then found to be polynomial functions of the combination 
$ \Nbar_h \xibar_h $, or $ \bbar \Nbar_h \xibar_h / \bbar_h $, 
\beqa
\Nbar_g \xibar_g ~~ &=& {\nbar_g \over \nbar_h} {\bbar \over \bbar_h} 
 \left( {\bbar \, \Nbar_h \xibar_h \over \bbar_h} \, 
  + {\bbar_h \bbar_2 \mubar_2 \over \bbar^2} \right) , \label{Nxi2} \\
\noalign{\smallskip}
\Nbar_g^2 \xibar_{3,g} &=& {\nbar_g^2 \over \nbar_h^2} 
 {\bbar \, \over \bbar_h}  \left[ S_{3,h} \, 
 \Bigl( {\bbar \Nbar_h \xibar_h \over \bbar_h} \Bigr)^2 
+ {3 \bbar_2 \mubar_2 \over \bbar} 
  {\bbar \Nbar_h \xibar_h \over \bbar_h} 
+ {\bbar_h \bbar_3 \mubar_3 \over \bbar^2} \right] ,  \nonumber \\  \\
\Nbar_g^3 \xibar_{4,g} &=& {\nbar_g^3 \over \nbar_h^3} \, 
{\bbar \, \over \bbar_h} \left[ S_{4,h} \, 
\Bigl( { \bbar \Nbar_h \xibar_h \over \bbar_h} \Bigr)^3 
 + {6 \bbar_2 \mubar_2 S_{3,h} \over \bbar} \, 
 \Bigl( {\bbar \Nbar_h \xibar_h \over \bbar_h} \Bigr)^2  \right. \nonumber \\
&& \qquad \left. {} + \Bigl({4\bbar_3\mubar_3 \over \bbar} 
+ {3 \bbar_2^2 \mubar_2^2 \over \bbar^2} \Bigr) \, 
 {\bbar \Nbar_h \xibar_h \over \bbar_h} 
 + {\bbar_h \bbar_4 \mubar_4 \over \bbar^2} \right] , \\
\noalign{\smallskip}
\Nbar_g^4 \xibar_{5,g} &=& {\nbar_g^4 \over \nbar_h^4} 
  {\bbar \, \over \bbar_h}  \left[ S_{5,h} \, 
 \Bigl( {\bbar \Nbar_h \xibar_h \over \bbar_h} \Bigr)^4 
 + {10 \bbar_2 \mubar_2 S_{4,h} \over \bbar} \, 
 \Bigl( {\bbar \Nbar_h \xibar_h \over \bbar_h} \Bigr)^3 
 \right. \nonumber \\
&& \qquad {} + \Bigl( {15 \bbar_2^2 \mubar_2^2 \over \bbar^2} 
  + {10 \bbar_3 \mubar_3 \over \bbar} \Bigr) \, S_{3,h} \,  
  \Bigr( {\bbar \Nbar_h \xibar_h \over \bbar_h} \Bigr)^2 \label {Nxi5} \\
&& \qquad \left. {} 
  + \Bigl( {10 \bbar_2 \bbar_3 \mubar_2 \mubar_3 \over \bbar^2} 
  + {5 \bbar_4 \mubar_4 \over \bbar} \Bigr)  
   {\bbar \Nbar_h \xibar_h \over \bbar_h} 
  + {\bbar_h \bbar_5 \mubar_5 \over \bbar^2} \right] , \nonumber 
\eeqa
etc.
For small $R$ the quantities $ \mubar_k $ rise monotonically with scale, 
but on large scales $ \mubar_k $ and $ b_k $ become constant \citep{FCFBST11}.

In the halo model, the galaxy correlations are not simply hierarchical, 
but every term in \eq{chi}, though no longer a simple power of 
$ \Nbar \xibar $, is a (polynomial) function of $ \Nbar_h \xibar_h $, 
If we assume that the halo distribution follows the hierarchical 
pattern $ \xibar_{k,h} = S_{k,h} \, \xibar_h^{k-1} $, 
then $ \chi $ is a function of the variable $ \Nbar_h \xibar_h $.
But, by equation (\ref{Nxi2}), $ \Nbar_g \xibar_g $ 
is a (linear) function of $ \Nbar_h \xibar_h $.
Thus, in the halo model, although the galaxy amplitudes $ S_k $ 
are not constant, $ \chi $ remains a function of $ \Nbar_g \xibar_g $: 
the hierarchical scaling for voids holds, even though 
$ \xibar_{k,g} $ no longer follows the simple hierarchical pattern.

The pattern is seen in general in the generating function formulation.
The probability generating function is additionally averaged 
over halo mass $m$, 
\beq
G(z) = \lexp g_{h}[g_{i}(z)] \rexp_m , 
\eeq
leading to the replacements of eqs.~(\ref{bmu_k}) and (\ref{bxi_k}); 
and so this pattern continues to all orders.
With no empty haloes, the halo occupancy generating function 
has $ g_i(0)|_m = p_0|_m = 0 $ for every halo mass $m$, 
and so we have the very useful result 
\beq
P_{0,g} = \lexp g_h \left[ g_i(0) \right] \rexp_m 
= \lexp g_h(0) \rexp_m = \lexp p_{0,h} \rexp_m = P_{0,h} . \label{P0gP0h}
\eeq
This is not a surprise: even when averaged over a distribution 
of haloes of different mass, 
no haloes means no galaxies, no galaxies means no haloes.

\section{Numerical Results}

We present results for statistics of voids in the distribution 
of dark matter, galaxies, and haloes for the numerical simulation 
studied in \citet{FCFBST11}.
The simulation is performed with the adaptive mesh refinement (AMR) 
code {\RAMSES} \citep{Teyssier02} for a $\Lambda$CDM cosmology 
with $\Omega_m=0.3$, $\Omega_\Lambda=0.7$, 
$ H_0 = 100 \, h \, \textrm{km} \, {\rm s}^{-1} \, \textrm{Mpc}^{-1} $ 
with $ h = 0.7 $, and normalization $\sigma_8=0.93$, where $\sigma_8$ 
is the root mean square initial density fluctuation in a sphere of 
radius $ 8 \Mpc $ extrapolated linearly to the present time.
The simulation contains $512^3$ dark matter particles on the AMR grid, 
initially regular of size $512^3$, in a periodic cube of size 
$ L_{\rm box}=200 \Mpc $; 
The hierarchical amplitudes $ S_k $ for mass, galaxies, 
and haloes in this simulation are presented by \cite{FCFBST11}, 
and further details can be found in \citet{CCT07}.

From the simulation data, 
we compute for spheres of radius $ R = 0.5 $--$ 25 \Mpc $ 
the probability $ P_0 $ that the volume be empty and the moments 
\beq
\Nbar = \lexp N \rexp , 
\eeq
and
\beq
\Nbar^2 \xibar = \lexp N^2 \rexp  - \lexp N \rexp^2 - \lexp N \rexp .
\eeq
The binomial uncertainty in the void probability is 
\citep{MLR87,H85} 
\beq
\Delta P_0 = \sqrt{ P_0 (1 - P_0) \over N_\tot } , 
\eeq
where $ N_\tot $ is the total number of independent volumes sampled, 
with is an additional cosmic variance contribution proportional 
to $ \xibar(L) $, the variance on the scale of the sample  
\citep{CBS95}, which is often insignificant.
In computing the uncertainty in $ \chi = -\ln P_0 / \Nbar $, 
the numerator and denominator are far from independent, 
but in fact are almost exactly anticorrelated, so that 
\beq
\Bigl( {\Delta\chi \over \chi} \Bigr) 
\approx \left| {\Delta P_0 \over P_0 \, |\log P_0| } 
 - {\Delta \Nbar \over \Nbar} \right| 
\eeq
\citep{CBS95}.
We adopt this as our error.

\begin{figure}
\includegraphics[width=\columnwidth]{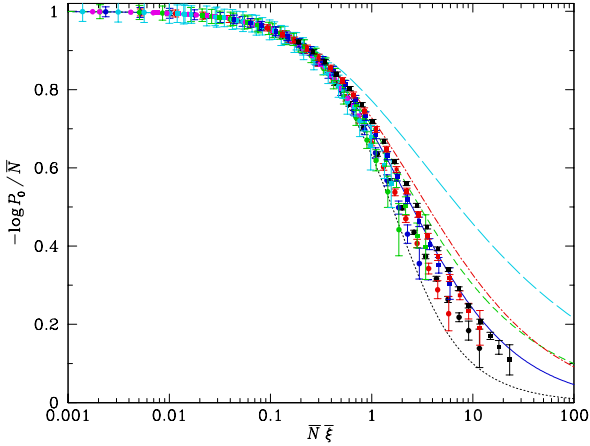}
\caption {Scaling statistic $ \chi = - \ln P_0/\Nbar $ 
plotted against scaling variable $ \Nbar \xibar $, 
for galaxies (squares) and haloes (circles).
Lines show models, as in Fig.~\ref{models} in the Appendix.
Colors red, blue, green, cyan, magenta show results 
for the full sample of haloes or substructure galaxies, 
and for random dilutions by successive factors of 2.
\label{v}}
\end{figure}

\begin{figure}
\includegraphics[width=\columnwidth]{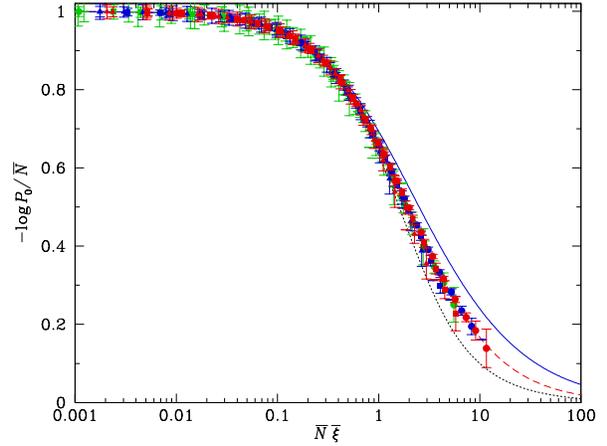}
\caption{Scaling curves for halo samples of different masses.
Red symbols show all haloes, 
blue symbols show haloes with mass $ M > 5 \times 10^{11} M_{\sun} $, 
and green symbols show $ M > 4 \times 10^{12} M_{\sun} $.
Circles, squares, and triangles show the full catalogs and 
dilutions by successive factors of two.
Dotted (black) and solid (blue) lines show the minimal and 
negative binomial curves, as before, and the dashed (red) line shows 
the geometric hierarchical model of \eq{chi_geom}.
\label{allhalos}}
\end{figure}

Figure \ref{v} shows the scaling behaviour of the void 
probability, $ -\log P_0/\Nbar = \chi (\Nbar\xibar) $, 
evaluated in the simulation.
Points represent results from spherical volumes of size $R$ 
ranging from $ R = 0.5 \Mpc $ to $ R = 25 \Mpc $.
Statistics are evaluated for the full substructure catalog, 
64\,316 substructures in 50\,234 haloes, 
and for random dilution by factors of 2, 4, and 8.
The two populations trace different, relatively well-separated loci, the 
upper points coming from the substructures and the lower from the haloes.
Curves show models as presented in Fig.~\ref{models} in the Appendix; 
the dotted (black) curve shows the minimal model (equation~\ref{chi_min}), 
the solid (blue) curve shows the negative binomial (equation~\ref{chi_nb}), 
the long dash-short dash (green) curve the quasi-equilibrium model 
of \citet{SH84}, the dot-dashed (red) curve the limiting lognormal 
or Schaeffer model (equation~\ref{chi_ln}), 
and the long-dashed curve the gravitational instability 
result of \citet{B92} before smoothing.
For small volumes,  $ \chi \to 1 $, the Poisson limit,
for all models, and the first correction $ 1 - \thalf \Nbar\xibar $
is also the same for all models; 
but for $ \Nbar \xibar \ga 1 $ differences begin to become apparent.
As found in observations, the substructure ``galaxies'' 
lie close to the negative binomial curve.
Figure~\ref{allhalos} shows the scaling behaviour of haloes of three 
different mass thresholds, from  $ 2 \times 10^{11} \, M_{\sun} $ 
to $ 4 \times 10^{12} \, M_{\sun} $.
Haloes of all masses are seen to follow well the middle curve, 
corresponding to the geometric hierarchical model of \eq{chi_geom}.

\begin{figure}
\begin{center}
\includegraphics[width=3in]{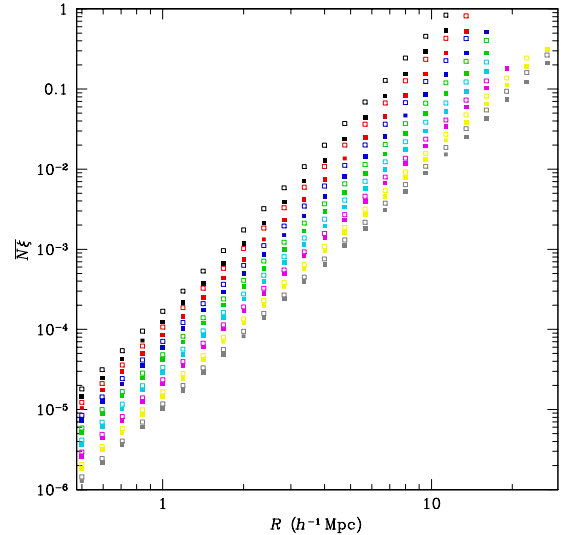}
\end{center}
\caption {$ \Nbar_g \xibar_g $ and $ \Nbar_h \xibar_h $ plotted 
vs. cell radius $r$.
Symbols of the same color show volumes of different radius 
for a fixed data sample.
Colors black, red, blue, green, cyan, magenta, yellow.
show random dilutions by successive factors of two, 
\label{fxxr}}
\end{figure}

\begin{figure}
\begin{center}
\includegraphics[width=3in]{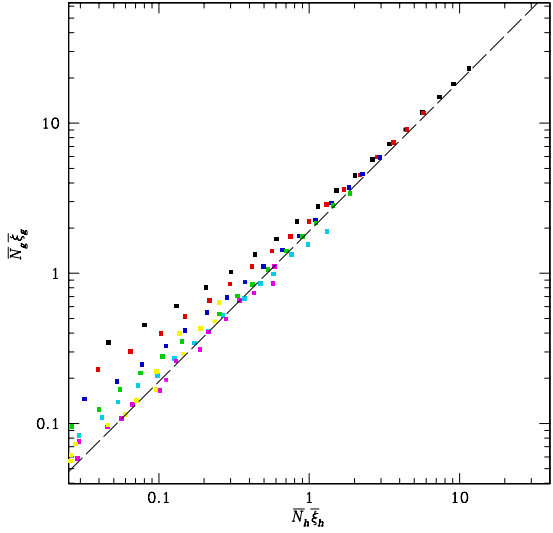}
\end{center}
\caption {$ \Nbar_g \xibar_g $ vs. $ \Nbar_h \xibar_h $ 
for the same data plotted in Fig. \ref{fxxr}.
Symbols of the same color show volumes of different radius 
for a fixed data sample: black shows the full sample; 
red, blue, green, cyan, magenta, and yellow.
show random dilutions by successive factors of two.
For colored symbols [red] etc., both galaxies and haloes 
are diluted by same factor.
\label{fxgxh}}
\end{figure}

\begin{figure}
\begin{center}
\includegraphics[width=3in]{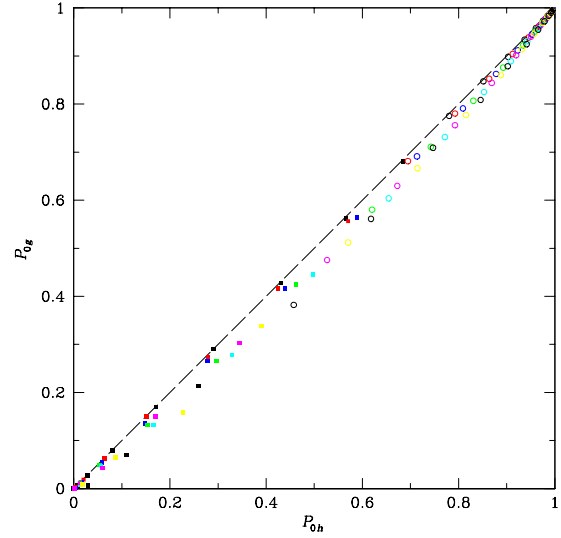}
\end{center}
\caption {Probability a volume is void of galaxies $ P_{0g} $ vs. 
probability the same volume is void of haloes $ P_{0h} $.
Symbols of the same color show volumes of different radius 
for a fixed data sample.
Different colors show random dilutions by a factor of two, 
black, red, blue, green, cyan, magenta, yellow.
\label{vgvh}}
\end{figure}

We see that the numerical results follow the predictions of 
hierarchical scaling, but this is not necessarily what is expected.
Within the uncertainties of sampling a small number of objects, 
the amplitudes for haloes may be consistent with constant values, 
but those for mass, and especially for galaxies, are not.
The normalization from $ \xibar_k $ to $ S_k = \xibar_k/\xibar_2^{k-1} $ 
removes much of the variation with scale (the unnormalized five-point 
function for mass covers more than ten decades), but the resulting 
$ S_k $ for galaxies are not constant, as shown in figs.~5 and 6 of 
\citet{FCFBST11}, 
where the residual variation is still a factor  of up to 10.
Thus, we are faced with the fact that hierarchical void scaling is 
observed, but its premise does not hold.
The halo model provides an explanation: in eqs.\,(\ref{Nxi2})--(\ref{Nxi5}), 
$ \xibar_2 $ and the higher order $ \xibar_k $ are all functions of 
$ \Nbar_h \xibar_h $, and $ \Nbar_h \xibar_h $ is linearly related to 
$ \Nbar_g \xibar_g $.
Figures~\ref{fxxr} and \ref{Nxi2} illustrate this in the simulation 
results.
Figure~\ref{fxxr} shows $ \Nbar_g \xibar_g $ and $ \Nbar_h \xibar_h $ 
as a function of scale $R$, for the full samples and 
for the two-, four-, an eight-fold dilutions.
Measured values are widely distributed; 
however, from \eq{Nxi2}, on large scales $\Nbar_g \xibar_g $ is 
related to $ \Nbar_h \xibar_h $, as illustrated in Fig.~\ref{fxgxh}.
On large scales, where halo size is negligible, 
we expect to have no galaxies only if we have no haloes, 
a result also implied in the composition of generating functions.
Figure~\ref{vgvh} shows $ P_{0,g} $ vs. $ P_{0,h} $
for the same volumes.

\begin{figure}
\includegraphics[width=\columnwidth]{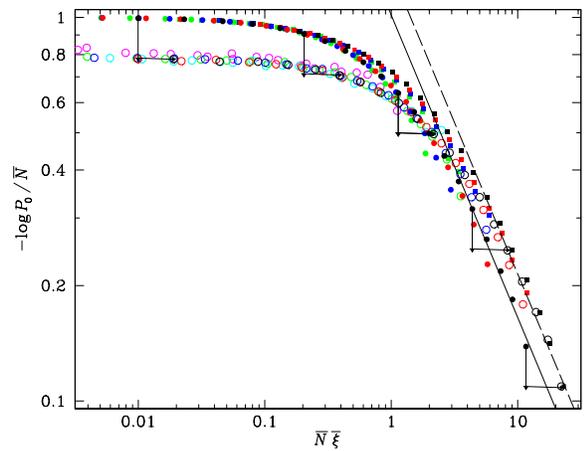}
\caption { 
Mappings between galaxy and halo scaling curves implied by the halo model.
Filled squares show the void scaling curve for galaxies, $ \chi_g = 
-\ln P_{0,g} / \Nbar_g $ as a function of $ x_g = \Nbar_g \xibar_g $, 
and filled circles show the scaling curve for haloes, $ \chi_h = 
-\ln P_{0,h} / \Nbar_h $ as a function of $ x_h = \Nbar_h \xibar_h $.
Open circles show the mapping of the halo data implied by the halo model, 
with $ \chi_h $ scaled by the factor $ \Nbar_h/\Nbar_g = 1/1.28 $, 
plotted as a function of $ x_h $ scaled by the factor 
$ 1.22^2 \times 1.28 = 1.91 $.
On large scales this becomes the same as the galaxy curve.
Straight lines show power-law behaviour $ x^{-\omega} $ 
with $ \omega = 0.79 $, separated vertically by a factor of 
$ (b_g/b_h)^{2\omega} \times (\Nbar_g/\Nbar_h)^{\omega-1} = 1.30 $ 
or horizontally by a factor 
$ (b_g/b_h)^2 (\Nbar_g/\Nbar_h)^{1-1/\omega} = 1.39 $.
\label{fpoint} 
}
\end{figure}

The halo model contains the requirement $ P_{0,g} = P_{0,h} $, 
no galaxies means no haloes, no haloes means no galaxies, 
from equation~(\ref{P0gP0h}) or from the fundamental sum over the 
occupancy of each halo, $ N_g = \sum N_i $.
From this, it is possible to obtain relations between void scaling 
curves for galaxies 
and their haloes, or between two different galaxy populations.
Figure~\ref{fpoint} illustrates a mapping suggested by the halo model.
from the halo curve to the galaxy curve.
The figure shows the scaling statistic for galaxies 
$ -\ln P_{0,g} / \Nbar_g $ (filled squares), for haloes 
$ -\ln P_{0,h} / \Nbar_h $ (filled circles), 
and for haloes mapped vertically by the ratio of number  
\beq
{\chi_g \over \chi_h} 
= {-\log P_0 /\Nbar_g \over -\log P_0/\Nbar_h} 
= {\Nbar_h \over \Nbar_g} 
= {1 \over \Nbar_i} = {1 \over 1.28} , \label{vertical}
\eeq
and horizontally by the ratio of the factors in $ \Nbar b^2 $, 
\beq
{\Nbar_g \xibar_g \over \Nbar_h \xibar_h}
= \Nbar_i \, (b_g/b_h)^2 = (1.28) (1.22)^2 = 1.91 \label{horizontal}
\eeq
(open circles).
The mapping is indicated by arrows for a selected sample of points, 
but every open circle originates from a filled circle.
The mapped halo curve and the galaxy curve are different for $ 
\Nbar \xibar \la 1 $, where resolved halo form factors affect the 
statistics \citep{FCFBST11}, 
but they merge for $ \Nbar \xibar \ga 1 $.
See also \citet{TC09}.

We obtain some inequalities comparing galaxies 
with their parent haloes.
Write the scaling variable as $ x = \Nbar \xibar $.
All halo scaling curves lie above the minimal scaling curve, 
and so $ \chi_h(x_h) > \chi_\min(x_h) > 1/x_h $.
With $ \chi_g/\chi_h = \Nbar_g/\Nbar_h $ and with 
$ x_g/x_h = b_g^2 \Nbar_g/ b_h^2 \Nbar_h $, 
this becomes a limit on $ \chi_g $, 
\beq
\chi_g = {\Nbar_h \over \Nbar_g} \, \chi_h 
= {(\Nbar_h b_h^2) \over (\Nbar_g b_g^2)} \, {b_g^2 \over b_h^2} \, 
 \chi_h = {x_h \over x_g} \, \chi_h {b_g^2 \over b_h^2} \, 
 > {b_g^2 \over b_h^2} \, {1 \over x_g} .
\eeq
Since $b(m)$ is an increasing function of mass and $ b_g $ 
(equation~\ref{bb}) is weighted to larger $m$ 
than $ b_h $ (equation~\ref{b_h}), we thus expect $ \chi_g > 1/x_g $.
As a horizontal scaling, we expect $ \chi_g = \chi_h $ at a value 
\beq
{x_g \over x_h} > {b_g^2 \Nbar_g \over b_h^2 \Nbar_h} .
\eeq
Since both $ b_g/b_h > 1 $ and $ \Nbar_g/\Nbar_h > 1 $, 
we expect that in general, galaxy scaling curve will be to the right 
of the halo scaling curve.

To the extent that the scaling curve can be represented as 
a power law, $ \chi = A x^{-\omega} $ \citep{BS88}, 
we can write quantitative relations.
Power-law behaviour implies a vertical mapping between 
two scaling curves at the same value of $x$, 
\beq
{\chi_g \over \chi_h} = \Bigl( {b^2_g \over b^2_h} \Bigr)^\omega
\Bigl( {\nbar_g \over \nbar_h} \Bigr)^{\omega-1} , 
\eeq
or a horizontal mapping, 
\beq
{x_g \over x_h} = {b^2_g \over b^2_h} \, 
\Bigl( {\nbar_g \over \nbar_h} \Bigr)^{1-1/\omega} 
\eeq
between two curves at the same value of $\chi$.
Since $ \omega $ is often near 1, the horizontal mapping is typically 
much more dependent on relative bias and only weakly on relative number.
This horizontal mapping is illustrated in Fig.~\ref{fpoint}, 
with $ \omega = 0.79 $.

We can compare two galaxy populations, say $i$ and $j$.
The simpler case is when both derive from essentially 
the same halo population; then we have the mappings 
\beqa
{\chi_i \over \chi_j} &=& {\nbar_j \over \nbar_i} , \\
\noalign{\smallskip}
{x_i \over x_j} &=& {b_i^2 \over b_j^2} \, { \nbar_i \over \nbar_j} .
\eeqa
With a power-law halo scaling function $ \chi_h = A x^{-\omega} $, 
we find the horizontal mapping $ \chi_i = \chi_j $ at 
\beq
{x_i \over x_j} = {b_i^2 \over b_j^2} \, 
\left( {\nbar_i \over \nbar_j} \right)^{1-1/\omega} .
\eeq
If halo scaling follows the minimal model, with $ \omega \approx 1 $, 
number density does not enter at all.
For the negative binomial model, on scales of interest 
$ \omega \approx 0.8 $ and $ 1-1/\omega \approx -0.25 $, 
still a very weak dependence.
Such a weak dependence on number means that we can expect scaling curves 
for populations with higher bias to be shifted to the right, 
by a factor of approximately $ b_{\rm rel}^2 $, 
or shifted upwards, by a slightly larger factor.

We can also compare galaxy populations that derive from distinct 
halo populations that have different number density and correlation 
strength, as long as they follow the same scaling curves, 
as in Figure~\ref{allhalos}.
Here, it is the relative bias $ b_{g,i}/b_{h,i} $ 
and relative number density $ \nbar_{g,i} / \nbar_{h,i} $, etc., 
that appear in the scaling relation.
For power law $ \chi \sim x^{-\omega} $, 
this reduces to the single horizontal scaling 
\beq
{x_i \over x_j} = {(b_g/b_h)_i^2 \over (b_g/b_h)_j^2} \, 
\left[ {(\nbar_g/\nbar_h)_i \over (\nbar_g/\nbar_h)_j} \right]^{1-1/\omega} .
\label{bgbhomega}
\eeq

\begin{figure}
\includegraphics[width=\columnwidth]{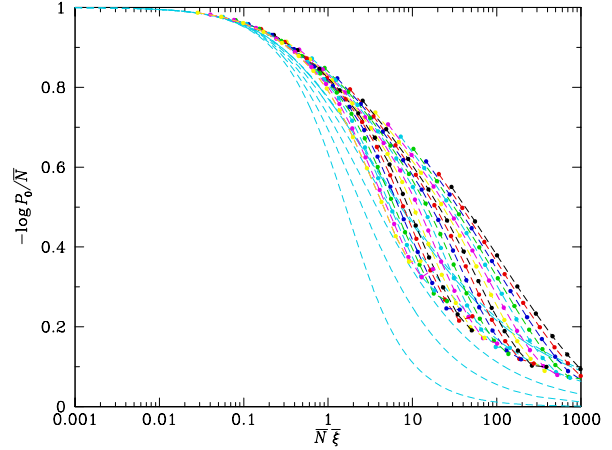}
\caption{Scaling function $ \chi = - \ln P_0/\Nbar $ plotted against 
scaling variable $ \Nbar \xibar $ for subsets of dark matter particles.
The top (black) curve is derived from a subsample of 
$ 256^3 = 16\,777\,312$ points, and each curve below that is 
diluted by a further factor of $\sqrt2 $; the final curve, 
diluted by a factor of 1024, then represents $16\,384$ points.
Dashed (cyan) curves show the behaviour expected from gravitational 
instability (equation \ref{sec_gi}), smoothed for spectral 
index $ n = +1 $, $0$, $-1$, $-2$, and $-3$ (bottom to top).
\label{voidDM}}
\end{figure}

Finally, we present results for voids in the mass 
or dark matter distribution.
The number of dark matter particles is so large that unless 
diluted substantially, only very small volumes are empty.
We begin with a random sample of one out of eight, or $256^3$ particles, 
for which we compute the full $P_N$ for volumes with 
$R = 0.2 \Mpc $ to $ R = 25 \Mpc $ by factors of $\sqrt2 $.
We then take advantage of the generating function to plot 
results for dilutions by a factor of $\lambda$, 
\beq
P_0(\lambda) = G(1-\lambda), 
\eeq
for $ \lambda = 2^{k/2} $, $k = 0$ to 20, 
or effective number of points from $ 256^3 = 16\,777\,216 $ 
down to $ 16\,284$.
Figure~\ref{voidDM} shows the scaling function 
$ \chi = -\ln P_0 / \Nbar $ plotted against the scaling 
variable $ \Nbar \xibar $ for the full sample and the twenty dilutions.
We note that \citet{2dF04} do not test scaling, but present results for 
only one density, which from their simulation parameters 
should be equivalent to the second curve below the median 
in Fig.~\ref{voidDM}.

\begin{figure}
\includegraphics[width=\columnwidth]{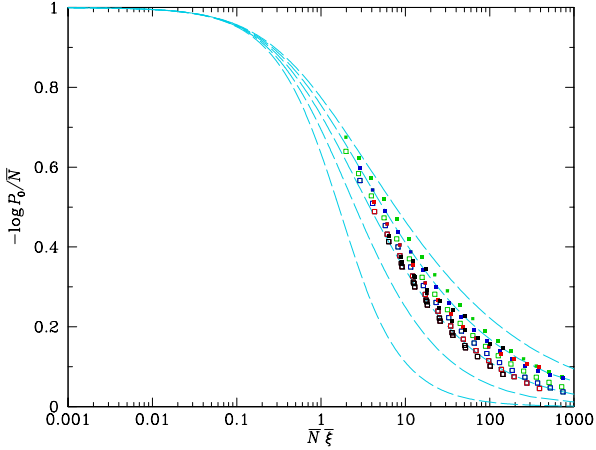}
\caption{Scaling curves $ \chi = - \ln P_0/\Nbar $ vs. 
$ \Nbar \xibar $ for point distributions that track dark matter, 
for large volumes: $ R = 6.3 $ (green), $ R = 8.8 $ (blue), 
$ R = 12.5 $ (red), and $ R > 16 \Mpc $ (black).
Filled symbols show measured results; open symbols show 
gravitational instability results smoothed for effective spectral 
index $n$, where $ 3+n = - \d\,\ln \xibar / \d\,\ln R $.
For these scales, $n$ takes on values $ -2 < n < -1 $.
Dashed (cyan) curves are as in Fig.~\ref{voidDM}.
\label{voidDMRlarge}}
\end{figure}

The dark matter results do not follow the scaling implied 
by gravitational instability, 
but this is because most of the volumes sampled are not in the 
large-scale, perturbative regime.
To compare with perturbative gravitational instability, 
Fig.~\ref{voidDMRlarge} shows results restricted to large 
volumes, $ R = 6.3 $, $8.8$, 12.5, and $ R > 17  \Mpc $.
For sufficiently large $R$ the measurements seem to 
approach the curve predicted for gravitational instability 
for the appropriate value of $ n \approx -2 $, 
where $ \d(\ln \xibar) / \d(\ln R) = - (3+n) $.

\begin{figure}
\includegraphics[width=\columnwidth]{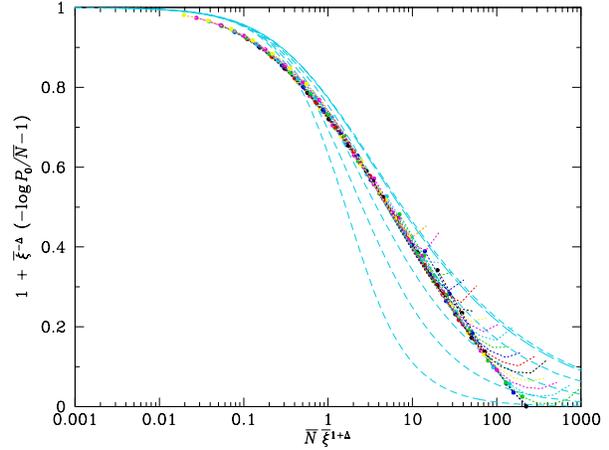}
\caption{The curves of Fig.~\protect{\ref{voidDM}} scaled 
as in \eq{Delta}, with $ \Delta = -0.0625 $.
Points are plotted for $ R < 1 \Mpc $, 
while connecting lines are extended for all $R$.
Dashed (cyan) curves are as in Fig.~\ref{voidDM}; 
measured values typically lie between $-2$ and $-1$. 
\label{voidDMRsmall}}
\end{figure}

The small-$R$ behaviour of the halo model also has its own, 
modified scaling behaviour, with power-law correlations 
$ \xibar_k \propto R^{-(k-1) \gamma + \delta } $,
where $ \gamma = (9+3n)/(5+n) $ and $ \delta = (3+n) p'/(5+n) $;  
$ p'$ characterizes the small-mass behaviour of the halo mass function, 
$ \d n/\d m \sim \nu^{p'}/m^2 $ \citep{MF00a,SSHJ01}.
In terms of $ \xibar_2 $, this is 
\beq
\xibar_k \sim \xibar^{(k-1) (1+\Delta) - \Delta} , 
\eeq
with $ \Delta = \delta/\gamma = p'/(3-p') $, 
independent of spectral index $n$.
This dependence implies a modified scaling, 
\beq
-\ln P_0/\Nbar = 1 + \xibar^{-\Delta} \; 
 \psi( \Nbar \xibar^{1+\Delta} ) ,  \label{Delta}
\eeq
expected to hold at small $R$.
This behaviour was anticipated numerically by \citet{CBH96}, 
who also point out that this modified scaling cannot 
persist on all scales.
Figure~\ref{voidDMRsmall} shows the success of this scaling 
for $ p' = -0.2 $, $ \Delta = -0.0625 $.
This value is different, even in sign, from the scaling exponent 
inferred from low order hierarchical amplitudes, 
although both are numerically small, 
and may indicate a change in the mass dependence of 
halo mass function at smaller masses.
In evaluating results for dark matter particles, we must 
also keep in mind that it is possible that some effect remains 
of the initial grid.
\citet{CBS95} suggest that void results should be reliable only 
$ P_0 \ga 1/e $, but we see no change of behaviour on 
different sides of this boundary.

\section{Discussion}

The implication of the halo model for void probability 
on large scales is simple: to have no galaxies means no halos, 
no halos means no galaxies.
This point cluster limit of the halo model provides a 
natural answer to the otherwise puzzling question, 
Why do voids obey the hierarchical scaling when the 
correlation functions do not satisfy the hierarchical premise, 
constant $ S_n $.
In the limit that the volume considered is large compared to 
halo sizes, void of galaxies means void of haloes, 
the void probability $ P_0 $ is the same, and scaling curves 
are related by number and by clustering strength or bias.
Since $ \Nbar_g > \Nbar_h $ (a halo contains one or more galaxies) 
and $ \xibar_g > \xibar_h $ (bias is an increasing function of mass), 
we anticipate points on the scaling plot move down and to the left.

In the simulations as well as in observations, the negative binomial 
scaling curve is a good approximation to that for galaxies; 
the weak clustering lognormal curve is much less favored, 
and the more strongly clustered lognormal even less so.
We expect the void scaling relation to provide different scaling curves 
for different galaxy populations; 
that galaxy results are often in agreement with the negative binomial 
curve can be attributed to the number density and clustering strength 
of typical galaxies.
With different results for different galaxy populations, 
in a direction consistent with relative bias and number density, 
we conclude that there is no fundamental reason that galaxies 
follow the negative binomial scaling curve, 
but that this follows from typical galaxy parameters.

The success of void scaling for galaxies requires that the underlying halo 
distribution follows the hierarchical pattern of higher order clustering.
In our simulations \citep{FCFBST11}, the halo $ S_{k,h} $ are 
approximately constant, roughly $ S_{3,h} \approx 1 $, 
$ S_{4,h} \approx 2 $, and $ S_{5,h} \approx 3 $, 
over a limited range of scales squeezed between 
the finite size of the simulation on the large end and the ability 
to separate extended objects on the small end.
These values of the $ S_{k,h} $ do not change much for different mass ranges.
More important, as seen in Fig.~\ref{allhalos}, 
different halo samples have remarkably similar scaling curves: 
for mass thresholds ranging over a factor of 20 
and number densities different by a factor of more than 4, 
the scaling curves are indistinguishable and seem to follow 
well the geometric halo mode curve of \eq{chi_geom}.
The scaling is important, because there is essentially no direct 
observational information for $ S_{k,h} $.
What results do exist are only for much higher mass thresholds: 
\citet{J90} measures the void scaling function for ACO clusters 
and \citet{CML91} for samples defined by \citet{PGH86} and \citet{T87}, 
but their results only reach $ \Nbar \xibar \la 2 $, 
for which all model scaling curves are much the same.
\citet{JZ89} find that Abell clusters have a hierarchical three-point 
function with amplitude independent of richness class, and 
\citet{CM95} also find, to a degree, constant $ S_k $ amplitudes for 
Abell and ACO clusters (but with a systematic difference between 
northern and southern galactic hemispheres), 
with numerical values $ S_3 \approx 3 $, $ S_4 \approx 15 $, 
$ S_5 \approx 100 $, appropriate to the high threshold, rare halo limit 
$ S_k = k^{k-2} $ of \citet{BS99}.
These do not apply directly to statistics of halos that host galaxies, 
including single galaxies and so extend down to galaxy masses; 
a theory that predicts the halo amplitudes $ S_{k,h} $ or 
the halo scaling curve for mass thresholds of $ 10^{11} $ or 
$ 10^{12} \, M_\odot $ has yet to be found.

Dark matter behaves differently.
For dark matter the behaviour of voids depends strongly on the 
density of particles.
In the quasilinear regime on large scales, the behavior seems 
to follow the predictions of gravitational instability.
For mass, there is no smallest object, no smallest cluster, and for 
any scale there are always clusters smaller and larger than that size.
The halo model has implications for high order functions, 
Small scales follow a modified scaling predicted by the halo model, 
as in equation~(\ref{Delta}).

\begin{figure}
\includegraphics[width=\columnwidth]{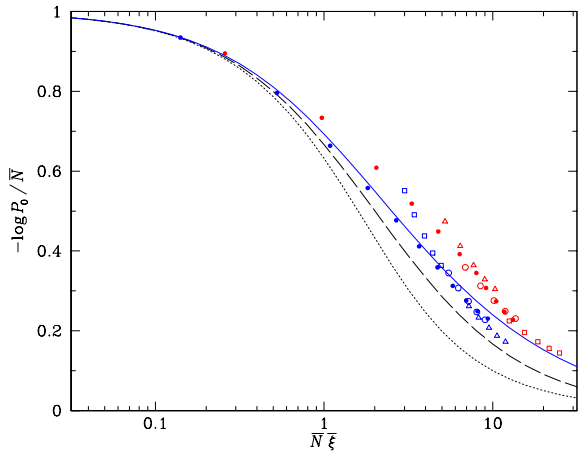}
\caption{Void scaling for SDSS red and blue galaxies.
Solid circles show SDSS data from \citet{TCNPW08}.
Open triangles show direct mapping, 
appropriate if the two samples inhabit the same haloes.
Open squares show mapping assuming the two samples derive 
not from the same haloes but from haloes that follow the 
same scaling curve.
Open circles are mapped only by bias, as appropriate for a 
power-law scaling with $ \omega = 1 $.
\label{tinker}}
\end{figure}

Halo model mappings have been derived in order to apply to observational data.
\citet{TCNPW08} present in their fig.~7($d$) void scaling curves 
for SDSS blue and red galaxies in which the locus for red galaxies 
is shifted substantially to larger values of $ \Nbar \xibar $; 
similar results are found for red and blue 2dFGRS galaxies 
by \citet{CNGB07}.
Figure~\ref{tinker} shows the SDSS data (filled circles) 
and the results of halo model scalings applied to red and blue galaxies 
for three different assumptions about the underlying haloes: 
assuming red and blue galaxies reside in the same haloes (open triangles), 
assuming a power-law halo scaling curve with $ \omega = 1 $ (open circles), 
and assuming their parent haloes trace same halo scaling curve (open squares), 
using the values $ b_\red = 1.02 $, $ b_\blue = 0.85 $, 
$ n_\red = 0.00328 $, $ n_\blue = 0.00433 \, h^3 \, \textrm{Mpc}^{-3} $, 
and ratios $ (b_g/b_h)_\red = 1.53 $, $ (n_g/n_h)_\red = 1.93 $, 
$ (b_g/b_h)_\blue = 1.18 $, $ (n_g/n_h)_\blue = 1.21 $.
computed from analytic halo occupation distributions 
for red and blue samples given by \citet{TCNPW08}.
Blue squares and red triangles begin to show departures from 
simple scalings, which should apply only in the large-scale limit.
The last, relative scaling is perhaps the most realistic, but the 
halo assumptions overlap and all of the scalings behave similarly.
This is confirmation that the ideas of the halo model apply to 
observations, as well as to simulations.

The void scaling results illustrate yet another success of the halo model, 
in describing nonlinear phenomena that it was not designed and 
not optimised to explain.
Applied to dark matter, the model may still be at best an approximation, 
but for galaxies, on scales where details of the structure of haloes are 
irrelevant, it is almost necessarily true: the total number of galaxies 
is the sum over haloes of the number of galaxies in each halo, 
and the combinatoric results of the halo model are independent of 
whether there is such a thing as a universal profile shape or not.

\null

\begin{center}Acknowledgments\end{center}

The question of why is it that voids in the galaxy distribution 
obey hierarchical scaling when their correlations do not 
follow the hierarchical pattern was raised by Darren Croton 
at the Summer 2007 workshop on Modelling Galaxy Clustering at 
the Aspen Center for Physics.
We thank David Weinberg for providing the SDSS results.
JNF acknowledges support from the City of Paris, Research in Paris 
program and thanks the Pauli Institute for Theoretical Physics, 
University of Z\"{u}rich, and the Institut d'Astrophysique de Paris 
for hospitality during this work.
This research has made use of NASA's Astrophysics Data System.


\def\aap{A\&A}
\def\aj{AJ}
\def\apj{ApJ}
\def\apjl{ApJ}
\def\apjs{ApJS}
\def\mnras{MNRAS}
\def\prd{PRD}

\appendix

\section{Models}

\begin{figure}
\includegraphics[width=\columnwidth]{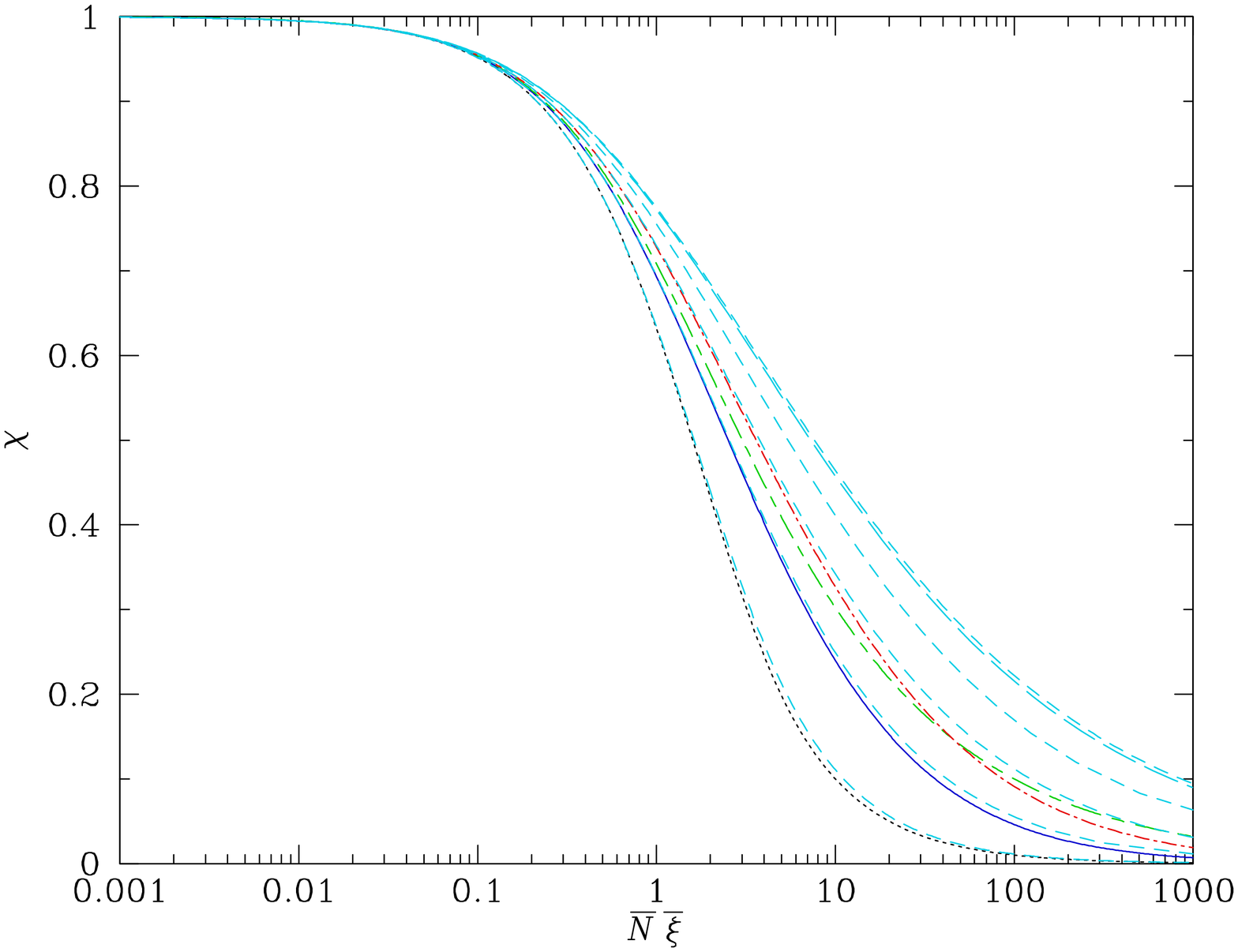}
\caption {Model void scaling functions $ \chi (\Nbar \xibar) $.
The solid (blue) line shows the negative binomial model 
(equation~\ref{chi_nb}); 
the doted (black) line shows the minimal model (equation~\ref{chi_min}); 
the short dash/long dash (green) line shows the quasi-equilibrium model 
(equation~\ref{chi_qe}); 
the dot-dash (red) line shows the Schaeffer, or lognormal model 
(equation~\ref{chi_ln}); 
the long-dashed (cyan) curve shows the gravitational instability 
prediction (Sec.\,\ref{sec_gi}); 
and the short-dashed (cyan) curves show the 
smoothed gravitational instability result for effective 
power index $n = -3 $, $-2$, $-1$, 0, and $+1 $ (top to bottom).
\label{models}}
\end{figure}

In this Appendix we present several models with specific 
analytic forms for the void scaling function $ \chi(\Nbar\xibar) $, 
useful against which to compare observational and numerical results.
Many of these models, introduced previously in a variety of different 
contexts \citep[see][]{F86,M07}, can be realised as halo models with 
a Poisson halo distribution; several were discussed by \citet{S96}.
With mean $\mu$, probabilities $ p_n = \mu^n \, e^{-\mu}/n! $, 
the Poisson generating function is 
\beq
g(z) = \sum_{n=0}^\infty {1 \over n!} \, \mu^n \, e^{-\mu} \, z^n 
= e^{\mu (z-1)} ; 
\eeq
in particular, the void probability is 
$ P_0 = e^{-\Nbar_h} = e^{-\Nbar/\Nbar_i} $.
For an unclustered halo distribution, correlations of galaxy number 
are given by the last term in eqs.~(\ref{xi2})--(\ref{xi5}), 
and a Poisson halo distribution is always hierarchical of a sort, 
with scaling function $ \chi = - \ln P_0/ \Nbar = 1/\Nbar_i $, 
scaling variable $ \Nbar \xibar = \Nbar_i \, \mubar_{2,i} $, 
and amplitudes $ S_k = \mubar_{k,i} / \mubar_{2,i}^{k-1} $ 
all determined by the occupation distribution 
(although not every occupation distribution has constant $ S_k $).
Fig.~\ref{models} compares models detailed in the following.

\subsection{Minimal Poisson Model}

A Poisson sum of clusters with mean $\mu$ 
with Poisson occupancy distribution with mean $\nu$ has 
\beq
G = g_h[g_i(z)] = \exp \left( \mu \, [ e^{\nu(z-1)} - 1 ] \right) , 
\eeq
From derivatives of $G$ we have moments 
$ \lexp N^{[k]} \rexp $ (equation~\ref{m_k}), 
\beqa
& \Nbar &= G'(1) = \mu \nu , \label{G'} \\ 
\noalign{\smallskip}
& \Nbar^2 \xibar &= G''(1) - [G'(1)]^2 =  \mu \nu^2 ,  \label{G''} 
\eeqa
from which we obtain $ \nu = \Nbar \xibar $, $ \mu = 1/\xibar $, 
void probability 
$ P_0 = G(0) = \exp\left[ - (1 - e^{-\Nbar \xibar})/\xibar \right] $, 
and thus 
\beq
\chi = {1 - e^{-\Nbar \xibar} \over \Nbar \xibar} . \label{chi_min}
\eeq
This minimal hierarchical model, with $ S_k = 1 $ for all $k$, 
saturates the Schwarz inequality requirement that the hierarchical 
amplitudes obey $ S_{2m} \, S_{2n} \ge S_{m+n}^{\, 2} $ \citep{F86}.

A Poisson occupation distribution formally includes 
possibility empty haloes.
The same result can be achieved by excising empty haloes 
and rescaling \citep{FCFBST11}, 
so that the occupation generating function becomes 
\beq
g_i(z) = {e^{\nu(z-1)} - e^{-\nu} \over 1 - e^{-\nu}} .
\eeq
The remainder of the calculation is straightforward; 
although the relation between $\nu$ and $\Nbar_i$ changes, again 
$ \nu = \Nbar \xibar $ and $ \chi = (1 - e^{-\Nbar\xibar})/\Nbar\xibar $.

Equation~(\ref{chi_min}) is the limit $ a \to \infty $ of 
the hypergeometric model of \citet{M07}, which has 
\beq
\chi_a = {(1 + \Nbar\xibar/a)^{1-a} - 1 \over (1-a) \Nbar \xibar / a} .
\eeq
The minimal model scaling curve is plotted as the dotted (black) 
line in Fig.~\ref{models}.

\subsection{Negative Binomial Model}

The negative binomial distribution with mean $\Nbar$ and parameter $K$ 
(also called Pascal, if $K$ is an integer, 
or P\'{o}lya distribution if $K$ is real), has count probabilities
\beq
P_N = { (N+K-1)! \over N! \, (K-1)!}
{(\Nbar/K)^N \over  (1 + \Nbar/K)^{N+K} } .
\eeq
For $ K=1 $ this reduces to the Bose-Einstein distribution, 
and is sometimes also referred to as modified Bose-Einstein.
This distribution 
appears in the frequency of industrial accidents \citep{GY20},
the distribution of ancient meteorites found in China \citep{YXL83},
in quantum optics \citep{KS68}, 
and in the multiplicity of charged particles  
produced in high energy collisions \citep{CS83,CS87,C91} 
and cosmic ray showers \citep{TCS87}, 
as well as in large-scale structure 
\citep{NSS53,CS83,CM83,C91,G92,EG92,GY93}, 
where it is often found to be a good approximation to 
the observed scaling curve of galaxies.

The negative binomial can be realised as a Poisson sum of clusters 
with logarithmic occupation distribution \citep{S95}.
The halo and occupation generating functions are 
\beqa
g_i(z) &=& \sum_{n=1}^\infty {-1 \over \ln(1-p)} \, {p^n \over n} \, z^n 
= {\ln(1-pz) \over \ln(1-\,p\,)} , \\
G(z) &=& \exp \left( \mu \, \Bigl[ {\ln (1-pz) \over \ln(1-\,p\,)} 
 - 1 \Bigr] \right) .
\eeqa
The probability of a void is $ G(0) = e^{-\mu} $, and $ \chi = \mu/\Nbar $; 
it is only necessary to relate these to moments $ \Nbar $, $ \xibar $ 
obtained from $ G'(0) $ and $ G''(0) $ as in eqs.~(\ref{G'}), (\ref{G''}), 
\beqa
\Nbar = -{\mu p /(1-p) \over \ln(1-p)} , \qquad 
\Nbar^2 \xibar = -{\mu p^2 / (1-p)^2 \over \ln(1-p)} 
\eeqa
Then, 
\beq
\chi = {\mu \over \Nbar} = {1-p \over p} \, \ln(1-p) , 
\eeq
\beq
\Nbar \xibar = {p \over 1-p} , 
\eeq
\beq
\chi = {\mu \over \Nbar} = {1-p \over p} \, \ln(1-p) 
= { \ln(1 + \Nbar \xibar) \over \Nbar \xibar} . \label{chi_nb}
\eeq
It has been suggested that convergence of the logarithmic series 
defined by \eq{chi_hier} with $ S_k = (k-1)! $ limits $ \Nbar \xibar < 1 $; 
but the probability generating function formulation has no restriction.

Equation~(\ref{chi_nb}) is the limit $ a \to 1 $ of 
the hypergeometric model of \citet{M07}.
The negative binomial model scaling curve is plotted as the 
solid (blue) line in Fig.~\ref{models}.

\subsection{Geometric Hierarchical Model}

An occupancy distribution with probability $ p_n \propto p^n $ 
for $ n \ge 1 $ has occupancy generating function 
$ g_i(z) = z(1-p) / (1-pz) $, and 
\beqa
G(z) &=& \exp \left( \mu \, \Bigl[ {z(1-p) \over 1-pz} - 1 \Bigr] \right) .
\eeqa
From the first and second moments, 
$ \Nbar = \mu p/(1-p) $ and $ \Nbar^2 \xibar = 2 \mu p/(1-p)^2 $, 
we find $ p = \half \Nbar \xibar / (1 + \half \Nbar \xibar) $, and 
\beq
\chi = {\mu \over \Nbar} = 1-p = {1 \over 1 + \thalf \Nbar \xibar} .
\label{chi_geom}
\eeq
The geometric halo model is the case $ a = 2 $ 
of the hypergeometric model of \citet{M07}, 
the model of \citet{H88} with $ Q = \half $, 
and also the $ \omega = 1 $ instance of the form  
$ \chi = 1/(1 + \Nbar \xibar/2\omega)^\omega $ cited in \citet{ABS90}.
Although not plotted, the geometric model falls between 
the minimal and negative binomial curves in Fig.~\ref{models}.

\subsection{Quasi-Equilibrium Model}

\citet{SH84} apply thermodynamics to obtain a gravitational 
quasi-equilibrium distribution function.
The resulting distribution is once again a halo model, a Poisson sum 
of haloes, with Borel occupation distribution \citep{SS94,S96}, 
\beq
p_n = {1 \over n!} \, (nb)^{n-1} \, e^{-nb} , 
\eeq
and with total count probabilities 
\beq
P_N = {\Nbar (1-b) \over N!} \, 
\left[ \Nbar (1-b) + N b \right]^{N-1} \, e^{-\Nbar (1-b) - Nb} .
\eeq
The void probability is $ P_0 = e^{-\Nbar (1-b)} $, 
and as the second moment gives 
\beq
1 + \Nbar \xibar = {1 \over (1-b)^2} , 
\eeq
the scaling function is 
\beq
\chi = {1 \over \sqrt{1 + \Nbar \xibar}} . \label{chi_qe}
\eeq
Saslaw and Hamilton assume the functional form 
$ b(\nbar T^{-3}) = b_0 \nbar T^{-3} / (1 + b_0 \nbar T^{-3}) $ 
to interpolate between ideal gas ($ b \to 0 $)
and virialized ($ b \to 1 $) limits.
\citet{S95} shows that invoking instead the form  
$  b = 1 - \ln(1 + b_0 \nbar T^{-3}) / b_0 \nbar T^{-3} $, 
(which has the same limits), 
the negative binomial also arises as a quasi-equilibrium model.

The quasi-equilibrium model scaling curve is plotted as the 
long dash/short dash (green) line in Fig.~\ref{models}.
This model is also the $ \omega = \thalf $ instance of the form  
$ \chi = 1/(1 + \Nbar \xibar/2\omega)^\omega $ 
cited in \citet{ABS90}.

\subsection{Lognormal Model}

It has been found that in the limit of very high threshold, 
a clipped Gaussian field produces a distribution with $ Q_k = 1 $, 
$ S_k = k^{k-2} $ for all $k$ \citep{PW84,S88}, 
the $ \nu = 0 $ model of \citep{S84} 
and a result that holds in  the rare halo limit 
under some very general condition \citep{BS99}.
For this set of amplitudes the scaling function is 
written parametrically \citep{S84} as 
\beq
\chi = (1 + \thalf \, \tau) \, e^{-\tau} , \qquad 
y = \Nbar \xibar = \tau \, e^{\tau}  . \label{chi_ln}
\eeq
This also constitutes the lower envelope of the lognormal distribution, 
suggested by \citet{H34} and more recently considered by \citet{CJ91}; 
although the full lognormal distribution does not in general scale, 
lognormal voids approach this curve for $ \xibar \ll 1 $ 
(numerically found to hold for $ \xibar \la 1 $).

The Schaeffer model, or lower bound of the lognormal distribution, 
is plotted as the long dash/short dash (green) line in Fig.~\ref{models}.

\subsection{Gravitational Instability \label{sec_gi}}

The gravitational instability amplitudes $ S_k $ can be computed 
in perturbation theory, which gives $ S_3 = 34/7 $ 
\citep{P80}, $ S_4 = 60\,712/1\,312 $ \citep{F84}, etc.
The complete set of amplitudes can be obtained 
from a generating function \citep{B92}.
In particular, the function 
\beq
\varphi(y) = \sum_{p=2}^\infty {(-1)^p \over p!} \,  S_p y^p 
\eeq
is obtained as a transform of the 
vertex generating function $ \G(\tau)$ by 
\beq
\varphi(y) = y \G + \thalf \tau^2 , \qquad \tau = y \G' .
\eeq
with $ \chi (y)= 1 + \varphi/y $.
The function $ \G(\tau) $ is found parametrically, 
\beqa
\tau &=& {3 \over 5} \left[ {3 \over 4} 
(\sinh \theta - \theta) \right]^{2/3} \\
\G &=& {9 \over 2} { (\sinh \theta - \theta)^2 
\over (\cosh \theta - 1)^3 } - 1 , 
\eeqa
the same hypercycloid functions that describe the 
time evolution of spherical underdensities 
\citep{P80}.
A useful analytic approximation to this function 
has been found to be 
\beq
\G = (1 + 2 \tau/3)^{-3/2} - 1 . \label{G32}
\eeq
The gravitational instability scaling curve is plotted as the 
long-dashed (cyan) line in Fig.~\ref{models}.

The smoothing in computing volume-averaged moments modifies 
the values of the $S_k$ and so also the scaling curve.
For a power-law power spectrum, $ P = A k^n $, 
\citet{B94} shows that the windowed vertex generating function 
becomes $\G^s = \G[ \tau (1 + \G^s)^{-(3+n)/6}] $.
With the approximation of \eq{G32}, the effect of smoothing 
on a scale where $ \xibar(R) $ has effective power index 
$ \d(\ln \xibar) / \d(\ln R) = - (3+n) $ then follows from 
\beq
\tau = {3 \over 2} \left( 1+\G \right)^{(3+n)/6} 
\left[ \left( 1 + \G \right)^{-2/3} - 1 \right] , 
\eeq
which can in some cases be solved analytically 
and in all cases can be used to obtain $\G$, $\varphi$, 
and $\chi$ numerically.
Dashed (cyan) curves in Fig.~\ref{models} show the 
windowed gravitational instability result for 
$ n = -3 $, $-2$, $-1$, 0, and $+1$ (top to bottom).
The $ n = +1 $ windowing of the gravitational instability 
scaling function is remarkably similar to the minimal model, 
and the $ n = 0 $ mapping of the gravitational instability function 
is remarkably similar to the negative binomial model.


\begin{thebibliography}{}


\bibitem[Alimi et al.(1990)]{ABS90} 
Alimi J.-M., Blanchard A., \& Schaeffer R., 1990, \apjl, 349, L5 

\bibitem[Balian \& Schaeffer(1988)]{BS88}
Balian R., Schaeffer R., 1988, \apjl, 335, L43 

\bibitem[Bernardeau(1992)]{B92}
Bernardeau F., 1992, \apj, 392 , 1

\bibitem[Bernardeau(1994)]{B94}
Bernardeau F., 1994, \aap, 291, 697

\bibitem[Bernardeau \& Schaeffer(1999)]{BS99}
Bernardeau F., Schaeffer R., 1999, \aap, 349, 697

\bibitem[Bouchet et al.(1993)]{BSDFYH93}
Bouchet F. R., Strauss M. A., Davis M, Fisher K. B., Yahil A., 
Huchra J. P., 1993, \apj, 417, 36

\bibitem[Cappi \& Maurogordato(1995)]{CM95}
Cappi A., Maurogordato S., 1995, \apj, 438, 507 

\bibitem[Cappi et al.(1991)]{CML91}
Cappi A., Maurogordato S., Lachieze-Rey M., 1991, \hfill\break
\aap, 243, 28 

\bibitem[Carruthers \& Minh(1983)]{CM83}
Carruthers P., Minh D.-V., 1983, Phys. Lett. 131B, 116

\bibitem[Carruthers \& Shih(1983)]{CS83}
Carruthers P., Shih C. C., 1983, Phys. Lett 127B, 242

\bibitem[Carruthers \& Shih(1987)]{CS87}
Carruthers P., Shih C. C., 1987, Internat. J. Mod. Phys. \hfill\break
A2, 1447

\bibitem[Carruthers(1991)]{C91}
Carruthers P., 1991, ApJ 380, 24

\bibitem[Coles \& Jones(1991)]{CJ91}
Coles P., Jones B., 1991, \mnras, 248, 1 

\bibitem[Colombi, Bouchet, \& Hernquist(1996)]{CBH96} 
Colombi S., Bouchet F.~R., Hernquist L., 1996, ApJ, 465, 14 

\bibitem[Colombi, Bouchet, \& Schaeffer(1995)]{CBS95}
Colombi S., Bouchet F. R.,, Schaeffer R., 1995, \apjs, 96, 401

\bibitem[Colombi, Chodorowski \& Teyssier(2007)]{CCT07}
Colombi S., Chodorowski M.~J., Teyssier, R., 2007, MNRAS, 375, 348 

\bibitem[Conroy et al.(2005)]{CCWNYCGDK05}
Conroy C. et al., 2005, \apj, 635, 990

\bibitem[Croton et al.(2004a)]{2dF04}
Croton D.~J. et al., 2004a, \mnras, 352, 828    

\bibitem[Croton et al.(2004b)]{Cetal04}
Croton D.~J. et al., 2004b, \mnras, 352, 1232   

\bibitem[Croton et al.(2007)]{CNGB07} 
Croton D.~J., Norberg P., Gazta{\~n}aga E., Baugh C.~M., \hfill \break
2007, \mnras, 379, 1562 

\bibitem[Elizalde \& E. Gazta\~naga(1992)]{EG92}
Elizalde E., Gazta\~naga E., 1992, MNRAS 254, 247

\bibitem[Fall et al.(1976)]{FGJW76}
Fall S. M., Geller M., Jones B. J. T., White S. D. M., 1976,
\apj, 205, L121


\bibitem[Fry(1984)]{F84}
Fry J. N., 1984, \apj, 279, 499

\bibitem[Fry(1986)]{F86}
Fry J.~N., 1986, \apj, 306, 358

\bibitem[Fry et al.(2011)]{FCFBST11}
Fry J. N., Colombi S., Fosalba P., Balaraman A., Szapudi I. 
Teyssier, R.  2011, \mnras, 415, 153

\bibitem[Fry et al.(1989)]{FGHMS89}
Fry J. N., Giovanelli R., Haynes M. P., Melott A. L., Scherrer R. J. 
1989, \apj, 340, 11

\bibitem[Gazta\~naga(1992)]{G92}
Gazta\~naga E, 1992, ApJ 398, L17.

\bibitem[Gazta\~{n}aga(1994)]{GAPM94}
Gazta\~{n}aga E., 1994, \mnras, 268, 913 

\bibitem[Gazta\~{n}aga \& Yokohama(1993)]{GY93}
Gazta\~{n}aga E., Yokohama J., 1993, \apj, 403, 450

\bibitem[Ghosh et al.(2001)]{GDGCKAS01}
Ghosh D., Deb A., Ghosh J., Chattopadhyay R., Kayum Jafri A., 
Azizar Rahman, M., Sarkar S.~R. 2001, Astroparticle Physics, 15, 329 

\bibitem[Greenwood \& Yale(1920)]{GY20}
Greenwood M., Yale G. U., 1920, J. R. Stat. Soc. A83, 255

\bibitem[Hamilton(1985)]{H85}
Hamilton A.~J.~S., 1985, \apjl, 292, L35 

\bibitem[Hamilton(1988)]{H88}
Hamilton A.~J.~S., 1988, \apj, 332, 67 

\bibitem[Hegyi(1992)]{H92} 
Hegyi S., 1992, Physics Letters B, 274, 214 

\bibitem[Hubble(1934)]{H34}
Hubble E., ApJ, 79, 8 (1934)

\bibitem[Jing(1990)]{J90}
Jing Y.-P., 1990, \aap, 233, 309

\bibitem[Jing \& Zhang(1989)]{JZ89}
Jing Y.-P., Zhang J.-L., 1989, \apj, 342, 639 

\bibitem[Klauder \&. Sudarshan(1968)]{KS68}
Klauder J. R., Sudarshan E. C. G., Fundamentals of Quantum Optics 
(Benjamin, New York, 1968).

\bibitem[Ma \& Fry(2000a)]{MF00a}
Ma C.-P., Fry J. N., 2000a, ApJ, 538, L107

\bibitem[Ma \& Fry(2000b)]{MF00b}
Ma C.-P., Fry J. N., 2000b, ApJ, 543, 503

\bibitem[Malik(1996)]{M96}
Malik S., 1996, \prd, 54, 3655 

\bibitem[Maurogordato \& Lachi\`{e}ze-Rey(1987)]{MLR87}
Maurogordato S., Lachi\`{e}ze-Rey M., 1987, \apj, 320, 13

\bibitem[Maurogordato et al.(1992)]{MSdC92} 
Maurogordato S., Schaeffer R., da Costa L.~N., 1992, \apj, 390, 17 

\bibitem[Mekjian(2007)]{M07}
Mekjian A.~Z., 2007, \apj, 655, 1 

\bibitem[Mo et al.(1997)]{MJW97}
Mo H.~J., Jing Y.~P., White S.~D.~M., 1997, \mnras, 284, 189 

\bibitem[Neyman \& Scott(1952)]{NS52}
Neyman J., \& Scott E.~L., 1952, \apj, 116, 144 

\bibitem[Neyman et al.(1953)]{NSS53}
Neyman J., Scott E.~L., Shane C.~D., 1953, \apj, 117, 92

\bibitem[Peebles(1980)]{P80}
Peebles P. J. E., 1980, The Large-Scale Structure of the Universe.
Princeton Univ Press, Princeton, NJ


\bibitem[Politzer \& Wise(1984)]{PW84}
Politzer H.~D., Wise M.~B., 1984, \apjl, 285, L1

\bibitem[Postman et al.(1986)]{PGH86}
Postman M., Geller M.~J., Huchra J.~P., 1986, \aj, 91, 1267

\bibitem[Ross et al.(2006)]{RBM06}
Ross A.~J., Brunner R.~J., Myers A.~D., 2006, \apj, 649, 48 

\bibitem[Ross et al.(2007)]{RBM07}
Ross A.~J., Brunner R.~J., Myers A.~D., 2007, \apj, 665, 67 

\bibitem[Saslaw \& Hamilton(1984)]{SH84}
Saslaw W. C., Hamilton A. J. H., 1984, \apj, 276, 13

\bibitem[Schaeffer(1984)]{S84}
Schaeffer R., 1984, \aap, 134, L15 

\bibitem[Scoccimarro, Sheth, Hui, \& Jain(2001)]{SSHJ01}
Scoccimarro R., Sheth R.~K., Hui L., Jain B., 2001, ApJ, 546, 20

\bibitem[Sheth(1995)]{S95}
Sheth R.~K., 1995, \mnras, 274, 213

\bibitem[Sheth(1996)]{S96}
Sheth R.~K., 1996, \mnras, 281, 1124 

\bibitem[Sheth \& Saslaw(1994)]{SS94}
Sheth R.~K., Saslaw W.~C., 1994, ApJ, 437, 35

\bibitem[Szapudi \& Szalay(1993)]{SS93} 
Szapudi I., Szalay A.~S., 1993, \apj, 414, 493 

\bibitem[Szalay(1988)]{S88}
Szalay A.~S., 1988, \apj, 333, 21 

\bibitem[Teich, Campos, \& Saleh(1987)]{TCS87}
Teich M.~C., Campos R.~A., Saleh B.~E.~A., 1987, PRD, 36, 2649

\bibitem[Teyssier(2002)]{Teyssier02}
Teyssier R., 2002, A\&A 385, 337

\bibitem[Tinker et al.(2008)]{TCNPW08}
Tinker J.~L., Conroy C., Norberg P., Patiri S.~G., 
 Weinberg D.~H., Warren M.~S., 2008, \apj, 686, 53

\bibitem[Tinker \& Conroy(2009)]{TC09}
Tinker J.~L., Conroy C., 2009, \apj, 691, 633 

\bibitem[Tully(1987)]{T87}
Tully R.~B., 1987, \apj, 323, 1 

\bibitem[Vogeley et al.(1994)]{VGH94} 
Vogeley M.~S., Geller M.~J., Park C., Huchra J.~P., 1994, 
\aj, 108, 745 

\bibitem[White(1979)]{W79}
White S. D. M., 1979, \mnras, 186, 145

\bibitem[Yang, Xuan, \& Li(1983)]{YXL83}
Yang W., Xuan R., Li L., 1983, Geochemistry, 1, 52

\end{thebibliography}
\end{document}